\documentclass[12pt]{iopart}
\usepackage{iopams}
\usepackage{graphicx}
\usepackage{epsfig}
\usepackage{cite}
\begin{document}

\title{The efficiencies of generating cluster states with weak non-linearities}

\author{Sebastien G R Louis$^{1,2}$,
Kae Nemoto$^{1}$, W J Munro$^{3,1}$, and T P Spiller$^{3}$}

\address{$^{1}$ National Institute of Informatics, 2-1-2
Hitotsubashi, Chiyoda-ku, Tokyo 101-8430, Japan}
\address{$^{2}$ Department of Informatics, School of Multidisciplinary Sciences,
The Graduate University for Advanced Studies, 2-1-2 Hitotsubashi, Chiyoda-ku, Tokyo 101-8430, Japan}
\address{$^{3}$ Hewlett-Packard Laboratories,
Filton Road, Stoke Gifford, Bristol, BS34 8QZ}
\ead{seblouis@nii.ac.jp}

\begin{abstract}
We propose a scalable approach to building cluster states of matter
qubits using coherent states of light. Recent work on the subject
relies on the use of single photonic qubits in the measurement
process. These schemes can be made robust to detector loss,
spontaneous emission and cavity mismatching but as a consequence the
overhead costs grow rapidly, in particular when considering single
photon loss. In contrast, our approach uses continuous variables and
highly efficient homodyne measurements. We present a two-qubit
scheme, with a simple bucket measurement system yielding an
entangling operation with success probability 1/2. Then we extend
this to a three-qubit interaction, increasing this probability to
3/4.  We discuss the important issues of the overhead cost and the
time scaling. This leads to a ``no-measurement" approach to
building cluster states, making use of geometric phases in phase space.
\end{abstract}


\maketitle

\section{Introduction}

The intriguing idea of one-way or cluster state quantum computing
was initially developed by Briegel and Raussendorf \cite{rau01}.
They showed that a two-dimensional array of qubits, entangled in a
particular way (through Conditional Phase gates), combined with
single qubit operations, feed forward and measurements are
sufficient for universal quantum computation. All the required
interactions are already contained inside the system, and the
computation proceeds through a series of local measurements (with
classical feed forward), efficiently simulating quantum circuits. In
effect, the logical gates are prepared off-line and imprinted onto
the qubits as they are transmitted through the cluster.

This approach was quickly applied
\cite{nie04,yor03,bro05,daw05,daw06} to linear optics quantum
computing \cite{KLM01} and was experimentally demonstrated on the scale of several qubits (see the review \cite{Kok06} for a full set of references). This scenario contains a significant
scaling problem in practice, due to the probabilistic nature of the
logical gates. However, the cluster state method enables this
problem to be pushed into the off-line preparation of the cluster
\cite{nie04,yor03,bro05}. Many different schemes have been proposed
to efficiently generate the photonic cluster states, because of the
simplicity of the interactions and the appealing coherence time of
the photons. Photon loss can be treated efficiently through
`indirect measurements' and a more elaborate preparation of the
cluster \cite{var05} but at a significant cost in terms of the qubit
dephasing \cite{rohde06}. There remains an issue concerning storage
though. Initially, each photon will be flying down an optical fibre
(or two \cite{yor03}), meaning there is a need for an adaptive
quantum memory. Reliable and efficient single photon sources and
detectors are a further issue for single photon approaches.

In the past few years, in an attempt to overcome the scaling
properties of linear optics quantum computing, new measurement-based
non-linear methods for optical quantum computing have been proposed
\cite{hut04,nem04,bar05,mun05a,mun05b,nem05,yamaguchi06}. One
approach is to effectively cascade non-demolition measurements, to
enable a parity check gate on two qubits \cite{nem04}. A pair of
photonic qubits prepared in a superposition of polarizations would
each in turn interact with the same coherent laser probe beam
$|\alpha\rangle$. The interaction consists of a cross-Kerr
non-linearity, which affects the phase of the probe beam, dependent
on the state of the individual photons. The weakness of the
non-linearity is compensated for by the size of $\alpha$. After
interacting with a pair of qubits, the probe beam undergoes a
homodyne measurement. The outcome of this measurement gives the
parity of the two qubits, thus projecting them into a known
entangled state heralded by the probe measurement result. By
combining two of these parity gates and an ancillary photon, a near
deterministic C-Not gate can be constructed \cite{nem04}, similar in
its form to the Pittman et al. C-Not gate in pure linear optics
\cite{pitfran}. However, the difference in the weak non-linearity
approach is that it is not destructive and not limited by the
beam-splitters' optimal success probability of $1/2$.

Already at this point, one can notice the usefulness of this parity
gate for photonic cluster state approaches. A near deterministic
entangler is all that is required to grow cluster states
efficiently, be it in Browne and Rudolph's \cite{bro05} or in Yoran
and Reznik's model \cite{yor03}. The required logical gates can be
obtained through entanglement and local operations alone. However,
as pointed out before, choosing photons as a support for one-way
quantum computing may not be the best option. Solid state or matter
systems may be more compact and easy to deal with in this
application and constituted the initial proposed system, when
cluster states were first developed \cite{rau01}. In many of the
solid state qubit systems proposed to date, the multi-qubit gates
arise from direct interactions between the qubits. Adding extra
qubits to a computation therefore leads to changes in the required
control fields and to the Hamiltonian of the whole system. As a
consequence, the required setup becomes increasingly more complicated as
the number of qubits in the computational system increases. A
further issue is that in order for some solid state qubits to
interact directly, they may need to be in such close proximity that
application of individual control fields and measurements becomes
infeasible.

To overcome these problems, the concept of distributed quantum
computing has arisen, in which interactions between the qubits used
in a computation is mediated by a third party. This third party can
be any measurable quantum physical system capable of interacting
with each qubit. Many proposals make use of single photons to
effectively mediate interactions between matter qubits
\cite{bos99,cab99,fen03,dua03,bro03,sim03}. Having interacted with
matter qubits, the photons then interact with each other in a linear
optical setup before being measured, thus projecting the matter
qubits into the required state without them having interacted
directly. It has been shown that entanglement and logical operations
can be generated in this way. However, once again there are
probabilistic limits in these approaches due to the fact that simple
linear optics is inherently non-deterministic.

The next step was to use these probabilistic entangling schemes to
prepare cluster states of matter
qubits \cite{bar05a,benjamin05,lim05,lim06,benjamin05a}. Barrett and
Kok first looked at this problem \cite{bar05a}, proposing the use of a
double-heralding probabilistic entangling procedure in order to
build cluster states. The mediators are single photons, mixed on a
beam splitter before being measured. The individual matter qubits
comprise of two low-lying (qubit basis) states and a single excited
state with a selection rule linking it to just one of the qubit
states. Applying a $\pi$ pulse leaves one of the low lying states
unchanged, while making the other move up to the excited state.
Decay leads to the emission of a single photon for this amplitude.
So if, after applying the $\pi$ pulse to both qubits, a single
photon is detected after the beam splitter, the qubits are projected
into the singlet state. The double-heralding procedure is used to
remove mixture, generated if non-photon-number-resolving detection
is used. This method has been further developed in a second paper,
using a repeat until success method proposed by Lim et al.
\cite{lim05}, where implementation of a conditional phase gate is
proposed, using a mutually unbiased basis \cite{lim06}. This enables
 some saving of qubit resources during the generation of the graph or
cluster states. However, a further very interesting aspect of this
proposal is that there are now three possible outcomes to the
measurement. Along with the usual success and failure outcomes,
there is an insurance outcome, in which the qubits are left in a
known product state, up to local operations. This means that,
following the insurance outcome, a new attempt to implement the gate
is possible. The corresponding scaling properties of the average
number of required entangling operations follow from the various
outcome probabilities for the entangling operation. This entangling
operation requires a rather elaborate measurement scheme, which may
be tough to implement experimentally. Furthermore, as the scheme
involves the detection of two photons, the success probability has a
quadratic dependence on the detector efficiency. Therefore on top of
the inherently probabilistic aspect of linear optics, the detector
efficiencies dramatically affect the scaling of the resources (even
for the highest reported efficiencies). It should be noted that the
scheme is robust against photon loss due to the fact that this is a
heralded source of error \footnote{This of course assumes no dark counts
in the detection process. Dark counts are generally an unheralded error
and unfortunately tend to be larger in the higher efficiency detectors.},
so the fidelity of successful operation
doesn't suffer. Nevertheless, the reduction in success probability
of the gate requires a significant increase in resource overhead,
which in turn increases the weight of unheralded errors in the
cluster state itself. So single photon measurement has its limits in
realizing entangling operations on matter qubits. However, homodyne
measurements on coherent light fields can be made much more
efficient than photon detection. In this paper we will show how this
and other factors make continuous variables a very powerful tool for
growing cluster states.

\section{The two-qubit entangling gate}

Recently, the weak non-linearity concepts \cite{nem04,mun05b} have
been applied to the area of matter qubits \cite{spi06} and the
distribution of entanglement \cite{loock06,ladd06}. There are quite
a number of well studied systems where one has a natural interaction
between a matter qubit and the electromagnetic field. These include
atoms (real and artifical) in cavity QED (both at the optical and telecom
wavelengths) \cite{raimond01}, NV-centers in diamond
\cite{jelezko04}, quantum dots with a single excess electron
\cite{pazy03}, trapped ions \cite{cir95} and SQUIDs
\cite{Shnirman97} to name only a few. All these systems are likely
to be suitable candidates for what we describe below but to
illustrate the details a little more clearly let us consider a
lambda based CQED system. One could use caesium atoms or an
NV-diamond centre embedded in the cavity. Both operate in the
optical frequency range and are in consequence well matched to
efficient homodyne measurements. The interaction between the
coherent field mode and our matter qubit can generally be described
by the Jaynes-Cummings interaction $\hbar g (\sigma^{-}a^{\dagger} +
\sigma^{+}a)$ and in the dispersive limit (large detunings) one
obtains an effective interaction Hamiltonian of the form
\cite{savage90,dispersive,dispersive2,bla04}:
\begin{equation}
 H_{int}=\hbar\chi\sigma_z a^{\dag} a.
\label{1}
\end{equation}
where $a$ ($a^{\dagger}$) refers to the annihilation (creation) operator
of an electromagnetic field mode in a cavity and the matter qubit is described
using the conventional Pauli operators, with the computational basis being
given by the eigenstates of $\sigma_z\equiv |0\rangle \langle 0|- |1\rangle \langle 1|$,
with $|0\rangle \equiv |\uparrow_z\rangle$ and $|1\rangle \equiv |\downarrow_z\rangle$.
The atom-light coupling strength is determined via the parameter $\chi=g^2/\Delta$,
where $2g$ is the vacuum Rabi splitting for the dipole transition and $\Delta$ is
the detuning between the dipole transition and the cavity field. The interaction
$H_{int}$ applied for a time $t$ generates a conditional phase-rotation $\pm\theta$
(with $\theta=\chi t$) on the field mode dependent upon the state of the
matter qubit. We call this a {\it conditional rotation} and it is very similar to the
cross-Kerr interaction between photons. This time dependent interaction
requires a pulsed probe.

Using this interaction, a two-qubit gate has been proposed \cite{spi06} based
on controlled bus rotation and subsequent measurement. The coherent state
$\alpha$ interacts with both qubits so an initial state of the system:
\begin{equation}
|\Psi_i\rangle=\frac{1}{2}(|00\rangle+|01\rangle+|10\rangle+|11\rangle)|\alpha\rangle
\label{2}
\end{equation}
evolves to
\begin{equation}
|\Psi_f\rangle=\frac{1}{2} \left(|00\rangle|\alpha
e^{2i\theta}\rangle+|11\rangle|\alpha e^{-2i\theta}\rangle +
(|01\rangle+|10\rangle)|\alpha\rangle\right). \label{3}
\end{equation}
At this stage we can choose from different types of measurements on
the probe beam. The first and simplest option we have is to perform
a homodyne measurement of some field quadrature
$X(\phi)=(a^{\dagger} e^{i \phi} + a e^{-i \phi})$ which for a
sufficiently strong local oscillator (compared to the signal
strength) implements a projective measurement $|x(\phi)\rangle
\langle x(\phi)|$ on the probe state \cite{tyc04}. The key
advantages with homodyne measurement, at least in the optical regime
are that it is highly efficient (99\% plus \cite{polzik92}) and is a
standard tool of continuous variable experimentalists. The easiest
to perform would be that of the momentum ($P = X(\pi/2)$)
quadrature. In that case the measurement probability distribution
has three peaks with the overlap error \cite{wjm05} between them
given by $P_{err}=\frac{1}{2}Erfc(\alpha\sin \theta /\sqrt{2})$. As
long as $\alpha \theta \sim \pi$ this overlap error is small
($<10^{-3}$) and the peaks are well separated. Now a measurement of
the central peak will project the two matter qubits into the
entangled state $(|01\rangle+|10\rangle)/\sqrt{2}$. This occurs with
a probability of 1/2. Detecting either of the other two side peaks
will project the qubits to the known product states $|00\rangle$ or
$|11\rangle$. It is worth explicitly mentioning here that we have
already reached the limits of conventional linear optics
implementations. When realistic detector efficiencies
($\eta\sim70\%$) are taken into account, the initial probability of
$1/2$ seen in photonic cluster states decreases proportional to
$\eta$ or $\eta^2$ depending on the specific implementation and so
the probability of the operation succeeding is now significantly
less than 1/2. However homodyne measurement are highly efficient and
so our success probability will remain very close to 1/2. It is of
course possible to exceed this 1/2 by changing the nature of our
measurement. In principle we could achieve a near deterministic gate
if we measured the the position ($X=X(0)$) instead of the momentum
($P$) quadrature . For the position quadrature there are two peaks
in the measurement result probability distribution, corresponding to
the even $|00\rangle+|11\rangle$ and the odd $|01\rangle+|10\rangle$
entangled states of the qubits. The overlap between the peaks
\cite{nem04} in this case is given by
$P_{err}=\frac{1}{2}Erfc(\alpha (1-\cos \theta) /\sqrt{2})$ which is
an issue because in order to separate the peaks well enough we would
require $\alpha \theta^2 \sim 2 \pi$. This is much more difficult to
achieve than $\alpha \theta \sim \pi$ required for the momentum
quadrature measurement.


Another strategy for the probe measurement would be to apply
an unconditional displacement $D(-\alpha)$ on the probe beam followed
by a photon number measurement. After the displacement the combined
state of the matter qubits and probe beam is
\begin{equation}
|\Psi_f\rangle
=\frac{1}{2}\left(|00\rangle|\alpha(e^{2i\theta}-1)\rangle+|11\rangle|\alpha(e^{-2i\theta}-1)\rangle
+ (|01\rangle+|10\rangle)|0\rangle\right) . \label{4}
\end{equation}
Now a photon number measurement of the bus mode will then either
pick out the vacuum state, or project onto two amplitudes
$\alpha(e^{\pm 2i\theta}-1)$ without distinguishing between them.
For an ideal projection onto the number basis $|n\rangle$, the state
of the two qubits becomes:
\begin{eqnarray}
\label{5}
|\Psi_f\rangle&=&\frac{1}{\sqrt{2}} (|01\rangle+|10\rangle)\qquad for\quad n=0 \\
\label{6}
 |\Psi_f\rangle&=&\frac{1}{\sqrt{2}} (|00\rangle+(-1)^n|11\rangle)\qquad for\quad n>0
\end{eqnarray}
with an equal probability of $1/2$ as long as the coherent amplitudes
$\alpha(e^{\pm 2i\theta}-1)$ do not contribute significantly to the vacuum.
The overlap of these coherent states $|\alpha(e^{\pm 2i\theta}-1)\rangle$ with
the vacuum $|0\rangle$ leads to an error probability of $P_{err}=e^{-4|\alpha\theta|^2}$
which can be made quite small with a suitable choice of $\alpha$ and
$\theta$ \cite{mun05b}. For example with $\theta$ small we can choose $\alpha\theta=2$
which leads to an error probability as low as $P_{err}\sim 3\times10^{-4}$.
Consequently we can obtain a near-deterministic gate if we can implement a photon number
measurement. However as observed in the linear optics schemes this currently
constitutes a significant technological challenge. In the future if this issue is
solved we will have a near deterministic entangling gate to build cluster states.
Without the photon number resolving detector, but assuming that the vacuum can be
distinguished from some photons our gate does work in a heralded fashion but with
a success probability of 1/2. 

Hence for cluster state generation the simplest
option so far would then be an efficient momentum ($P = X(\pi/2)$) quadrature homodyne
measurement, giving us a success probability close to $1/2$. We will note here that this is the most accessible and robust weak-nonlinear scheme so far proposed, using a single interaction per qubit. Would it be
possible to further improve the success probability all the
while maintaining a highly efficient measurement?

\section{The three-qubit entangling gate}

\begin{figure}[!htb]
\begin{center}\includegraphics[scale=0.6]{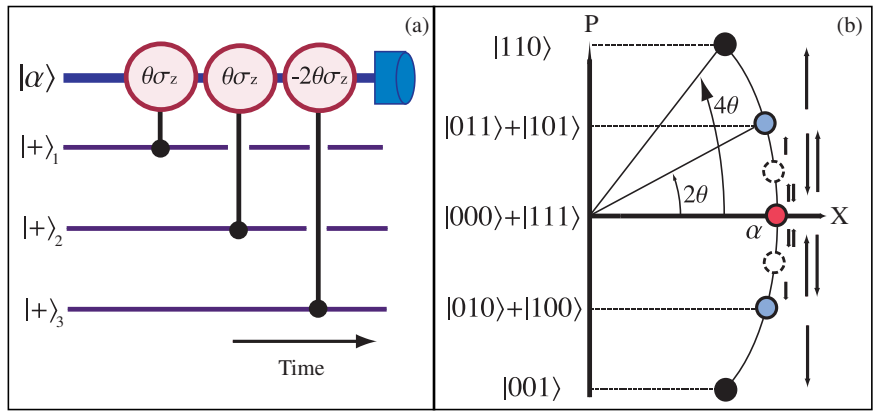}\end{center}
\caption{Schematic diagram (a) of a three qubit entangling
operation. In (b) the possible probe trajectories caused by the
three conditional rotations. There are five different end-states.
Upon measurement, three of these will project the qubits to
entangled states of interest.} \label{fig1}
\end{figure}

Within the same framework of conditional rotations, one can envisage
three qubits interacting with a single probe beam. If we limit
ourselves again to efficient $P$ quadrature measurements (which
scale as $\alpha \theta$), we could consider the generation of three
qubit states. GHZ states are for instance one particularly useful
state \cite{bro05}. One way of projecting the qubits onto GHZ-type
states is to vary the strength of the interactions between the
qubits and the probe beam. Let us represent a rotation of the
coherent probe beam by $R(\theta ) = \exp(i \theta a^{\dagger}a )$.
Now no $R(\pm\theta\sigma_{z_{1}}) R(\pm\theta\sigma_{z_{2}})
R(\pm\theta\sigma_{z_{3}})|\alpha\rangle$ combination will lead to
the required GHZ end states in the case that we implement a $P$
quadrature measurement. However having one of the qubits interact
twice as much with the probe beam will yield the correct paths in
phase space. Consider the sequence
$R(\theta\sigma_{z_{1}})R(\theta\sigma_{z_{2}})
R(-2\theta\sigma_{z_{3}})|\alpha\rangle$ which we depict in figure
\ref{fig1}.
The peak centered on the origin will then correspond to the GHZ
state $(|000\rangle+|111\rangle)\sqrt{2}$ (after being detected).
This will happen with a probability of 1/4. Next the two peaks
having been rotated through $\pm2\theta$ will correspond to the
qubit states $(|01\rangle_{1,2}+|10\rangle_{1,2})|1\rangle_{3}/\sqrt{2}$
and $(|01\rangle_{1,2}+|10\rangle_{1,2})|0\rangle_{3}/\sqrt{2}$
respectively. Now in both of these possible outcomes we obtain
the same Bell state on qubits 1 and 2, disentangled with qubit 3.
So overall we obtain a GHZ state with probability of 1/4 and a
Bell state with probability of 1/2 (on two qubits of our choice),
heralded by the probe beam $P$ quadrature measurement outcome.
The other two outcomes project the qubits to two different known
product states $|001\rangle$ or $|110\rangle$. Consequently, if all we
want to do is entangle a pair of qubits, we can now do this with
probability of 3/4 \footnote{It is worth noting that this success
probability is higher than that of two successive parity gates.}.

It may seem like increasing the number of qubits taking part will
further raise the success probability. This claim turns out to be
valid if we allow for more and more interactions as we add extra
qubits. Considering for example the 4 qubit case. The optimal
combination then becomes
$R(\theta\sigma_{z_{1}})R(\theta\sigma_{z_{2}})R(2\theta\sigma_{z_{3}})R(-4\theta\sigma_{z_{4}})|\alpha\rangle$.
We now have 16 possible paths in phase space with 9 different end
states. All of these apart from two, under detection, will project
the qubits to Bell states and GHZ states. Focusing solely on qubits
1 and 2 (these can be any two qubits which we choose to have
interact only once with the probe beam), they will be entangled with
probability $p=7/8$. Following this method for larger numbers of
qubits,
$R(\theta\sigma_{z_{1}})R(\theta\sigma_{z_{2}})R(2\theta\sigma_{z_{3}})R(4\theta\sigma_{z_{4}})...R(-2^{n-2}\theta\sigma_{z_{n}})|\alpha\rangle$
the success probability in entangling a specific pair of qubits
(here 1 and 2) scales as $p=1-2^{1-n}$. We don't necessarily have to
view these extra $n-2$ qubits as ancillas. They can become (if they
aren't already) useful elements (`dangling bonds') for future
operations when we consider the generation of 2-dimensional cluster
states. However there are drawbacks to using this generalization.
The setup and measurement process will become increasingly
complicated. The probe beam will have to travel and interact a lot
more, rapidly accentuating the errors that we could have had
initially. Another essential point to note, is that the gate
operation time will grow exponentially with the number of qubits we
are willing to use. If we only have access to a fixed interaction
strength $\theta$, the gate operation time will double every time we
add an extra qubit. So depending on the situation we are in, a
compromise will have to be made between the time we are willing to
take and the success probability we want to achieve. The 3-qubit gate minimizes the ratio of operation time over probability and we shall use this 3/4 probability in the remainder of the paper.

\section{Scaling}

We now consider how this entangling scheme may be used for
generating cluster states of matter qubits. The usual cluster state
is a rectangular 2-dimensional lattice of qubits. The qubits are
entangled in a particular way, through conditional phase gates, with
some of their nearest neighbours, thus creating `bonds'. Each
1-dimensional chain represents the life line of a single qubit to be
processed. These chains form a full 2-dimensional lattice structure
by having bonds between them. The cluster state is defined as the
eigenstate of the set of operators $S_i=X_i\prod_jZ_j$, where $i$
represents a particular lattice site and $j$ all its nearest
neighbours.

Building chains is a possible basis for generating cluster states.
If the chains can be efficiently generated, then simple schemes can
be developed to combine them to form a 2-dimensional cluster,
required for quantum computing \cite{nie04,bro05,dua05}. Given a
parity check operation, the simplest growing technique involves
taking one qubit (prepared in a superposition state
$(|0\rangle+|1\rangle)/\sqrt{2}$) at a time and linking it on the
end of the chain. Once this is done, a Hadamard transform is
performed on this new end qubit, before the next one is added. In
case of failure, the initial end qubit is left in an unknown state.
Thus it needs to be measured--adaptive feed forward on its nearest
neighbour then enables recovery of the cluster state. So the chain
shrinks by one qubit in this case. This constitutes the basic
sequential approach to building chains. Clearly for success
probabilities smaller than 1/2, the chain will shrink on average;
for a success probability of exactly 1/2, it will remain the same
length. We can immediately appreciate benefits from the relatively
high probabilities achieved in our two entangling procedures. The
first two-qubit procedure already constitutes the limit of simple
linear optics approaches. The second one, involving a 3 qubit
interaction, can already be used in a sequential fashion, ensuring
fast average growth and thus limited resource consumption.

\begin{figure}[!htb]
\begin{center}\includegraphics[scale=0.45]{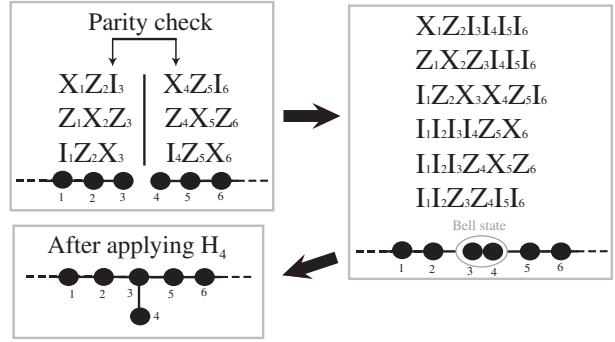}\end{center}
\caption{Applying a parity check projects the two involved qubits to a state
stabilized by the operator $Z_3Z_4$, removing all the operators
anti-commuting with it. Here we apply a Hadamard transform on
qubit 4 after the operation, thus producing a dangling bond.}
\label{fig2}
\end{figure}

\begin{figure}[!htb]
\begin{center}\includegraphics[scale=0.45]{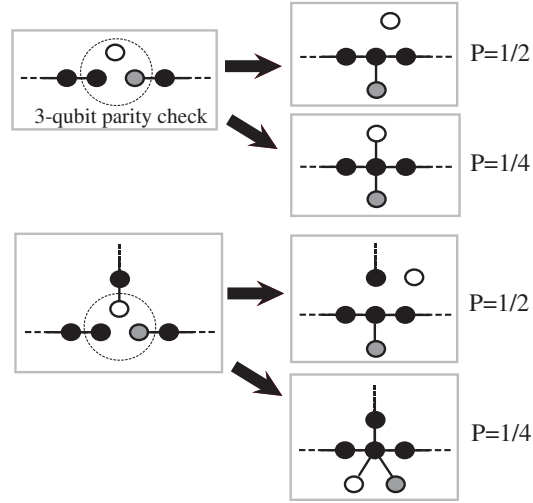}\end{center}
\caption{Using the 3-qubit gate, we can first attempt to join a pair of chains.
This will work 3/4 of the time, producing one or two dangling bonds centred on
the same qubit, allowing for repeated trials at linking chains to
form two dimensional clusters. Three chains can also been linked
up in a similar fashion to produce a `T' shape.}
\label{fig3}
\end{figure}

In the case of lower probabilities, small chains are to be built
inefficiently before joining them to the main chain. The process of
linking chains with an entangling operation is described in figure
\ref{fig2}, using the stabilizer notation. We can see that even
though we are not obliged to measure out one qubit, the actual
length of the resulting chain is the sum of the two initial ones
minus one qubit. A convenient way of representing this operation
with states is used in \cite{bar05a}. And in figure \ref{fig3}, we
can see how the three qubit gate allows us to directly join three
chains into a `T' piece, as well as joining two chains together. In
case of failure, the end qubits need to be measured out and each
chain shrinks by one element. Then the procedure is repeated.
Supposing we start off with two chains of equal length $L$
(following previous analysis \cite{dua05}, $L$ is defined as
the number of qubits constituting the linear cluster), then the
average size of the resulting chain is:
\begin{equation}
L_f=\sum^L _{i=0} 2(L-1/2-i)p(1-p)^i\approx 2L-1-2(1-p)/p \; .
\label{7}
\end{equation}
This approximation isn't meaningful for small chains. Here $p$ is the
success probability of the entangling operation. We can immediately identify
a critical length, above which there is growth:
\begin{equation}
L>L_c=1+2(1-p)/p \; . \label{8}
\end{equation}
This critical length varies between different entangling
operations. If an actual logical gate can be immediately
implemented, then $L_c=2(1-p)/p$ for example. Or if this logical
gate requires the qubits from the cluster to interact directly
(non-distributive approach) then $L_c=4(1-p)/p$ \cite{dua05}. So our minimal size chain, ensuring growth is $L_0$ and $L_0>L_c$
(the next integer greater than $L_c$).

Following strategies previously developped, these minimal chains are grown in a divide and conquer fashion (in parallel, without recycling) starting from individual qubits before being merged together. This yields scaling relations for the average time taken $T[L]$ and the average number of entangling operations $N[L]$ to grow a chain of length $L$:
\begin{eqnarray}
\label{10and11}
T[L]&=& t\sum_{i=1}^{\log_2(L_0-1)+1}(1/p)^i+(t/p)\log_2\left(\frac{L-L_c}{L_0-L_c}\right) \\
N[L]&=&\frac{\left(1/2\sum_{i=1}^{\log_2(L_0-1)+1}(2/p)^i+1/p\right)(L-L_c)}{L_0-L_c}-\frac{1}{p}
\end{eqnarray}
where $t$ denotes the time taken per entangling operation.

For our first entangling procedure $p=1/2$, $L_c=3$ and thus
$L_0=4$. Growing this 4-chain will require 14 entangling operations on
average leading to $N[L]=16L-50$. This is
already the theoretical limit for simple single photon
applications. In the repeat until success method \cite{lim06}, for a
failure probability of 0.6 (and equal success and insurance
probabilities, on all results), the scaling is $N[L]=185L-1115$
and for a failure probability of 0.4 it becomes
$N[L]\simeq16.6L-47.7$. Now if we switch to our 3 qubit gate, then
$L_0=2$. We will notice that $L_c=5/3$ meaning the $L_0-L_c$
factor will contribute more than before, because we chose this
difference to be unity (also note that here this difference tends
to unity as the success probability increases). The average number of entangling
operations required then simply becomes $N_0=1/p=4/3$, giving us a
scaling $N[L]=8L-44/3$. This is a vast improvement over previous proposals.

We shall extend this scaling comparison to the generation of
2-dimensional cluster states. Using the redundant encoding method
described in \cite{bro05}, we can give the average number of qubits
consumed in the creation of a vertical link. Each trial to
establish this link costs two qubits (per chain), because we first need to
create dangling bonds. If we succeed in linking the
two dangling bonds, we need to measure the first dangling qubit in
order to establish the C-phase gate then measure the next one, to
have a direct link between the two chains. The fact that we can
only implement a simple entangling operation and not the logical
gate means we lose an additional qubit, which may be used later
for additional vertical links or error correction. But if we
concentrate on the task of making a single vertical link, the
number of qubits consumed is:
\begin{equation}
V=2(1/p+1) \; . \label{12}
\end{equation}
We can see that this converges to 4 as p tends to unity
(this corresponds to the qubit cost of a single trial). Then the
average number of entangling operations required to make the
vertical bond is given by:
\begin{equation}
\label{13}
 N_V=2N[V]+1/p \; .
\end{equation}
Using the linear optics scheme proposed in \cite{lim06}, for failure
probabilities of 0.6, 0.4 and 0.2 respectively, $N_V=3334$, $191.2$
and $32.5$. The latter failure probability would however be very
difficult to implement physically. With our two and three qubit entangling procedures
we obtain $N_V=70$ and 46.7 respectively. We can see that the efficiency
of these schemes generalizes to the creation of 2-dimensional cluster
states in a straightforward manner. Our gates can also be used to build
cluster states in a `tree' like fashion, as proposed by Bodiya and Duan
in \cite{bod06}. The method relies on the observation that GHZ-type states
are locally equivalent to star shaped cluster states. A parity check is
all that is needed here.

We now turn back to the time scaling. Solving $T$ for $p=1/2$ we end up with:
\begin{equation}
\label{14}
 T[L]=14+2\log_2(L-3) \; .
\end{equation}
Of course this is only valid for $L\geq4=L_c$. The above result is
obtained with a $T_0=14$ corresponding to the average time needed
to build a 5-qubit chain without recycling (this is due to the
form of the sum). It is pretty clear that if we allow for entangling
operations to be made in parallel, alongside additional resources, this $T_0$
can be minimized. For $p>1/2$ we have $L_0=2$, meaning we only keep the first
term in the sum for $T_0$. This results in a general closed expression
for $T$:
\begin{equation}
\label{15}
 T[L]=(t/p)\left(1+\log_2\left(\frac{L-L_c}{L_0-L_c}\right)\right) \; .
\end{equation}
We can compare this with the time taken by a sequential adding and
building, as we now have access to probabilities higher than
$1/2$. Adding one qubit at a time, via an entangling procedure,
gives the recursion relation $L_{k+1}=L_k+2p-1$ for the length, leading
to the number of rounds $k=(L-1)/(2p-1)$. So for our 3 qubit gate, on
average the chain grows by one unit every two trials. The time being
sequential too, $T_{L+1}=T_L+t/p$, the general form for $T$
becomes:
\begin{equation}
\label{16}
 T[L]=(L-1)t/p \; .
\end{equation}
Thus time now scales linearly with the length of the chain in
contrast with the logarithmic dependence we had above.

\section{Optimizing time and resources}

For the two-qubit entangling gate, we essentially stand at the
same point as the photonic cluster state approaches. Optimizing
the resources boils down to finding the optimal strategies in
combining elements of cluster states. Though this is a classical
analysis, relying on probabilistic gates, it is a very complex task.
Obtaining bounds or comparing different strategies requires computing
assistance. In their recent paper, Kieling, Gross,  and Eisert \cite{kie06,gro06}
investigate these issues in significant detail. They analyse essential
methods and derive bounds for the globally optimal strategy,
based on an entangling operation working with probability 1/2.

For higher probabilities however, the critical length insuring average growth simply doesn't exist anymore and additional truly scalable approaches are at hand.
We shall go over the obvious ones. From previous works on generating cluster
states \cite{bar05a,dua05}, we know that the simplest way to grow short chains
with probabilistic gates is through a `divide and conquer' approach. It also turns
out to be much quicker than a sequential adding, as we allow for many gates to operate
in parallel. As described earlier on, this technique attempts to link up chains of
equal length on each round, and discards the chains which failed to do so.

This approach can be extended to growing large chains in the aim of saving time.
In this context we can work out some important \textit{average} quantities,
starting off with the time taken:
\begin{equation}
\label{17}
 T[L]=1+\log_2(L-1)=k \; .
\end{equation}
Here $k$ represents the number of rounds and can easily be worked out, as we saw above,
from the given chain length. Thus we will only use $k$ in the following expressions.
Next we can give the number of chains, at a particular round $k$ ($L=1$ for $k=0$),
having started off with $n$ qubits:
\begin{equation}
\label{18} C[k]=n(p/2)^k  \; .
\end{equation}
Then the number of remaining qubits on that round is given by:
\begin{equation}
\label{19}
 Q[k]=C[k]\times L=n(p/2)^k(2^{k-1}+1) \; .
\end{equation}
Following this we can work out the number of wasted qubits $W[k]=n-Q[k]$. Finally,
when discussing the necessary resources we need the overall cumulative number
of entangling operations:
\begin{equation}
\label{20}
G[k]=\sum_{j=1}^{k-1}\frac{C[j,m]}{2}=\frac{n}{2}\left(\frac{1-(p/2)^{k-1}}{2/p-1}\right)
\; .
\end{equation}

\begin{figure}
\begin{center}\includegraphics[scale=0.65]{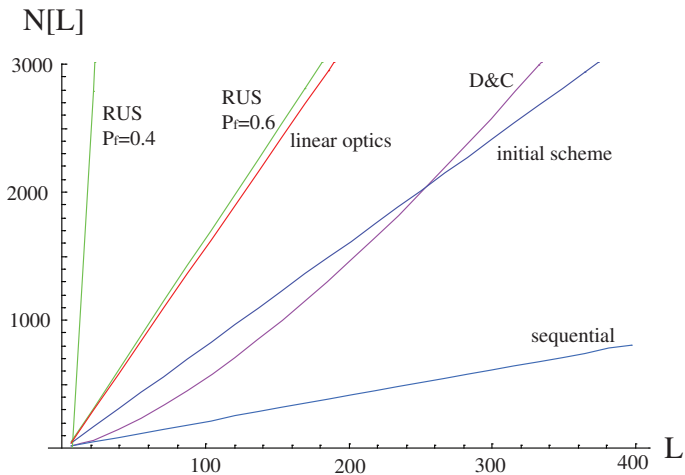}\end{center}
\caption{Comparison of entangling operation requirements for chain production, using our 3-qubit gate. Clearly for chains smaller than 250 qubits the full divide
and conquer approach is more advantageous than the linear scaling obtained through
the initial merging technique. The savings in the number of entangling operations
are most significant around lengths of 100 to 120 qubits. However, the sequential
adding scheme is significantly more efficient still, as expected. With this we achieve
much lower scalings in comparison with those obtained through the repeat until
success (RUS) scheme, $P_f$ representing the failure probability. The `linear optics' curve corresponds to a success probability of 1/2 using the divide and conquer strategies \cite{dua05}. This is the theoretical limit of conventional linear optical schemes, ignoring all detector and source inefficiencies.}
\label{fig4}
\end{figure}

In order to have a first comparison with the method described in the previous section, we
can set the value of $C[k]$ to unity. Or alternatively, we can use the ratio
$N_{dc}[k]=G[k]/C[k]$ which will give the average number of entangling operations
required to produce a single chain:
\begin{equation}
\label{21}
 N_{dc}[k]=\frac{\sum_{j=1}^{k-1}C[j]}{2C[k]} \; .
\end{equation}
Expressing this ratio in function of $L=2^{k-1}+1$ we obtain:
\begin{equation}
\label{22}
 N_{dc}[L]=\frac{(2/p)^{\log_2(L-1)}-1}{2-p} \; .
\end{equation}
From the initial strategy, with $m\geq2$ we reached a value linear in $L$:
\begin{equation}
\label{23}
 N[L]=\left(\frac{2}{p}\right)\frac{L-1-2(1-p)/p}{1-2(1-p)/p}-\frac{1}{p} \; .
\end{equation}
Obviously this will scale better with $L$, but surprisingly enough,
the threshold above which it becomes more advantageous is very high.
As observed in figure \ref{fig4} (for our 3 qubit gate), up till
lengths of 250 qubits, the full divide and conquer approach requires
less entangling operations. This is due to the fact that the
probabilities we are dealing with are significantly higher than in
previous schemes, which were undertaken in two steps, the building
of minimal elements $L_0$ and then their merging, in order to be
scalable.

We can compare this with the sequential adding method, as we now have access to probabilities
higher than $1/2$. Adding one qubit at a time, via an entangling procedure, gives the length's
recursion relation $L_{k+1}=L_k+2p-1$ leading to the number of rounds $k=(L-1)/(2p-1)$. For
our 3 qubit gate, on average the chain grows by one unit every two trials. In this case, the
number of rounds is equivalent to the number of entangling operations so we have:
\begin{equation}
\label{24}
 N_{seq}[L]=(L-1)/(2p-1) \; .
\end{equation}
Obviously this represents a considerable saving, as can be verified
in figure \ref{fig4}.

We can also compare the time scaling of these various strategies, in units of time
$t$ corresponding to a single measurement. For the complete divide and conquer scheme
we simply have:
\begin{equation}
\label{25} T_{dc}[L]=t\left(1+\log_2(L-1)\right) \; .
\end{equation}
and for the initial scheme:
\begin{equation}
\label{26}
T[L]=\frac{t}{p}\left(1+\log_2\left(\frac{L-L_c}{L_0-L_c}\right)\right)
\; .
\end{equation}
For the sequential adding, the cumulative time obeys $T_{L+1}=T_L+t/p$, and the
general form for $T$ becomes:
\begin{equation}
\label{27} T_{seq}[L]=t(L-1)/p \; .
\end{equation}
Thus time now scales linearly with the length of the chain, in contrast with
the logarithmic dependence we had above.

\begin{figure}
\begin{center}\includegraphics[scale=0.65]{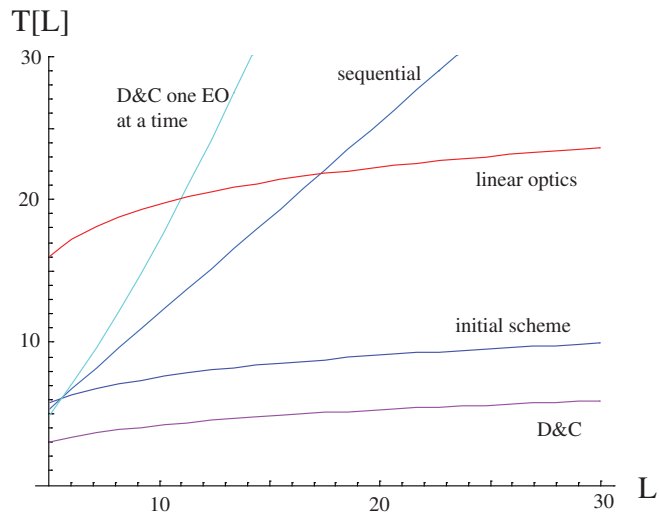}\end{center}
\caption{Comparison of time requirements for chain production, for various
approaches, as a function of the chain length. The divide and conquer approach,
as expected, saves significant amounts of time. The linear time dependence of
sequential adding does not compare, for long chains. However we can see the
difference in work required if we only allow one entangling operation at a time. Again we can observe the theoretical limit for linear optics.}
\label{fig5}
\end{figure}

The first two approaches have a logarithmic dependence on the length
$L$, however $T_{dc}$ is significantly lower as might have been
expected (see figure \ref{fig5}). Overall we see that there is a
clear advantage to divide the task up and to run parallel entangling
operations. However we also note that the resources in qubits become
quite large, in the absence of recycling. The amount of wasted
qubits for the full divide and conquer approach grows very quickly
as can be seen from the expression for $W[k]$. One could envisage in
this case a form of partial recycling, to save on the qubit
resources whilst still retaining the time speed up. Then we would
allow for two or three trials before discarding the chains (the
initial scheme set no limit on the number of trials). However the
protocol now becomes more elaborate unless we are willing to wait
between each round (of discarding) because some chains will link on
the first trial while others will link on the second (supposing we
allow two trials). So it seems like savings in time could be made if
we are able to manage and organize chains of different lengths.

The linear time scaling for the sequential method is due to the fact that operations cannot
be undertaken in parallel during its growth. If we didn't have access to simultaneous
entangling operations, the time scaling for the divide and conquer methods would be equivalent
to $N_{dc}[L]$ which is sub-exponential. One needs to keep in mind that by adopting a sequential
method, the whole procedure is simplified considerably and would be more accessible to physical
implementations. Divide and conquer methods require a lot of work in parallel and should in practice
involve the moving about and reordering of qubits or even small cluster states. Unless the actual
edges linking up the vertices in the graph states can be displaced via entanglement swapping
strategies, we will most probably have to physically move some vertices in order to implement
additional entangling operations. Adding qubits sequentially should solve some architectural
problems which may arise. For example, the qubits could be perfectly static and the measuring
system (including the ancillary qubit which can be reused) would travel along the chain `zipping'
it up. Of course the measuring system would go back and forth, with a frequency related to the
success probability of a single operation. But essentially the qubits constituting the chain
wouldn't have to move. This seems significantly more practical than moving the qubits and
chains around or having to change the measuring setup every time so as to implement the
operation between qubits in various places. However many of these problems
may be solved by more advanced protocols making use of percolation
phenomena as developed in \cite{perco}.

All this brings us to view the cluster state as having active
regions in which it is being built or measured in the computation
(both can be undertaken simultaneously) and regions in which the
qubits are simply waiting. Now this waiting can be minimized in the
building itself, through the appropriate protocols, and in the
measurement process. That is, the cluster can be built only a few
layers in advance, so that the qubits have less waiting to do,
between the building and the actual measurement. In any case, there
will be some waiting. Therefore the lowest decoherence support would
be preferred, but it may not be the easiest to manipulate. Thus we
may envisage having two different physical realizations constituting
the cluster state. For example, we could use single electron spins
initially in building the cluster. Once the links are made between
one site and its nearest neighbours, the qubit could be switched
into a nuclear spin state which has a significantly longer coherence
time, via a swap operation or some other coherent write and read
actions. Most of the waiting would be done in the long-lived state,
before being swapped again for the readout \cite{kan98,loock06}.
This follows the same scenario as using a second physical system to
mediate the interaction and make the measurements, in distributed
quantum computing. In the present proposal, we use a continuous
variable bus and homodyne measurements to generate the links. This
physical system shows itself to be very efficient in this
application. Then, for example, electron spins or superconducting
charge qubits could then be the matter realization interacting with
the bus and serving for the final readout. These systems provide the
useful static aspect required, they interact well with the mediating
bus and ensure good single qubit measurements. Finally a low
decoherence realization such as nuclear spin could be envisaged,
mainly as a storage medium. The swapping or write and read procedure
should have a high fidelity for this storage to be beneficial. On
the whole, we see that optimization will depend directly on the
physical realization(s) we have chosen to work with. For systems
with long dephasing times we would give priority to sequential
adding approaches, as we have some freedom in the time scaling and
thus we can make significant savings in resources. But for
realizations with short dephasing times, we would probably want to
divide the task up and run operations in parallel, in order to
accelerate the fabrication of the cluster state, at the expense of
extra resources.

\section{The measurement-free approach}

Looking at our entangling gates, we have seen that if we utilize
four non-linear interactions and three qubits the success probability is
dramatically increased. Within the framework of four non-linear interactions,
another option presents itself to us \cite{spi06}. Defining the conditional
displacement operator by $\hat D(\beta\sigma_z)=\exp\left(\left\{\beta\hat a^\dagger-\beta^*\hat a\right\}\sigma_z\right)$, one can simulate a conditional-phase gate between qubits 1 and 2 with the following interaction sequence:

\begin{equation}
\hat{U}=D(-\beta_{2}\sigma_{z_2})D(-\beta_{1}\sigma_{z_1})D(\beta_{2}\sigma_{z_2})D(\beta_{1}\sigma_{z_1})=e^{2iIm\left\{\beta_{1}^{\ast}\beta_{2}\right\}\sigma_{z_1}\sigma_{z_2}},
\label{28}
\end{equation}
by setting $\beta_{1}^{\ast}\beta_{2}=i\pi/8$. The resulting operation is then
locally equivalent to that of the conventional conditional phase
$e^{i\pi(1-\sigma_{z_1})(1-\sigma_{z_2})/4}$. Thus at the end of the
sequence the coherent state is disentangled from both qubits,
removing the need for a measurement. These conditional displacements
can be simulated through conditional rotations and \textit{unconditional}
displacements, as is clear from the self-inverse quality of Pauli operators
and the general property of rotations \cite{louisell}:

\begin{equation}
e^{\theta a^{\dag}a}f\left(a,a^{\dag}\right)e^{-\theta a^{\dag}a}=f\left(ae^{-\theta},a^{\dag}e^{\theta}\right),
\label{29}
\end{equation}
where $f$ can be expanded in a power series. This can be realized with the following sequence \cite{loock}

\begin{equation}
D(\alpha\cos\theta)R(-\theta\sigma_z)D(-2\alpha)R(\theta\sigma_z)D(\alpha\cos\theta)=D\left(2i \alpha \sin\theta\sigma_z\right),
\label{30}
\end{equation}
with $\alpha$ real.
The conditional phase observed arises from the
different areas the probe traces out in phase space, picking up a geometrical
phase. More in the spirit of the initial proposal of Wang and Zanardi \cite{zanardi}
this can easily be extended to the simulation of many-body interactions.
The interactions required to build a cluster state are pairwise thus conditional
displacements are sufficient.

\begin{figure}
\begin{center}\includegraphics[scale=0.55]{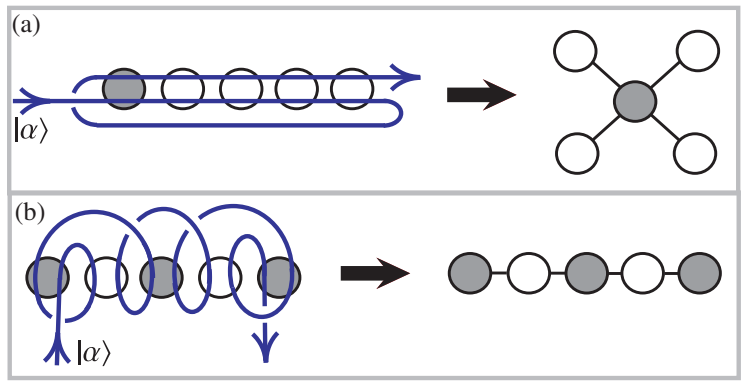}\end{center}
\caption{Schematic representation of two different interaction sequences derived in the text.
The gray and the white qubits lead to a pure imaginary and pure real conditional
displacement of the probe mode $\alpha$. The first and second interaction, for
each qubit, are in opposite directions in phase space. In (a) we have the sequence
leading to a star shape while in (b) the one leading to a linear graph, both on 5 qubits.
For larger numbers of qubits we have the same patterns.}
\label{fig6}
\end{figure}

By having the probe interact with more qubits and adapting the direction
of the displacements, we can pick out the qubits we want to couple through
the geometrical phase. In that way one could start from a general sequence of the form:

\begin{equation}
\!\!\!\!\!\!\! \prod_{n=1}^N D(-\beta_n\sigma_{z_n})\prod_{n=1}^N
D(\beta_n\sigma_{z_n})=\exp\left[2i\Im\left\{\sum_{n=1}^{N-1}\left(\beta_n^{\ast}\sigma_{z_n}\sum_{p=n+1}^N\beta_p\sigma_{z_p}\right)\right\}\right].
\label{31}
\end{equation}
But here we are simulating interactions between all qubits and from
this sequence one cannot directly generate linear or grid-like cluster states.
That is we need to adjust the $\beta_n$ so as to choose which qubits we want to interact. We can use such a sequence to
directly generate useful graph states such as star shaped graphs (locally equivalent
to GHZ states). For example, taking $\beta$ real and the displacement from qubit 1
orthogonal to all the others we have

\begin{equation}
\!\!\!\!\!\!\! \!\!\!\!\!\!\! \!\!\!\!\!\!\! \prod_{n=2}^N
D(-\beta\sigma_{z_n})D(-i\beta\sigma_{z_1})\prod_{n=2}^N
D(\beta\sigma_{z_n})D(i\beta\sigma_{z_1})=\exp\left[-2i\beta^2\sigma_{z_1}\left\{\sum_{n=2}^N\sigma_{z_n}\right\}\right].
\label{32}
\end{equation}
Clearly if we set $\beta=\sqrt{\pi/8}$ we will obtain a star shaped
graph of our $N$ qubits, centered on qubit 1 (figure \ref{fig6}(a)).
We note that from this condition on $\beta$ and equation~(\ref{30}),
the scaling and magnitude requirements for $\alpha \theta$ here are comparable
to those of the measurement-induced entangling schemes.
To generate a linear cluster we need to switch to another
interaction sequence. We need to disentangle the probe with the
qubits as the sequence evolves so as not to create extra links.
Coming back to our conditional phase  operation $\hat{U}$ we notice
that after the third interaction the probe becomes disentangled from
qubit 1. Furthermore by setting $\beta_{1}=i\beta_{2}$ the entirety of the
geometric phase is already acquired (by the corresponding two-qubit
state) at this point. Along with this observation and a correct
ordering of the displacements we can propose a sequence of the form

$$
D(-\beta\sigma_{z_N})D(-i\beta\sigma_{z_{N-1}})...D(\beta\sigma_{z_4})D(-\beta\sigma_{z_2})D(i\beta\sigma_{z_3})D(-i\beta\sigma_{z_1})D(\beta\sigma_{z_2})D(i\beta\sigma_{z_1})
$$
%
\begin{equation}
=\exp\left[2i\beta^2\sum_{n=1}^{N-1}(-1)^n\sigma_n\sigma_{n+1}\right]
\label{33}
\end{equation}
Again setting $\beta=\sqrt{\pi/8}$ all the couplings are locally
equivalent to conditional phase gates, yielding a linear cluster
state (see figure \ref{fig6}(b)). We can view the probe as creating
the links as it travels along the chain.

The main advantage with these generalizations is the reduced number of
interactions per qubit. If we were to use the simple conditional phase sequence
$\hat{U}$ then the number of interactions per qubit would be $2d$ where $d$
is the degree of the qubit in the graph state. In other words, to build a $N$
qubit star shaped graph, the center qubit would have to interact $2(N-1)$ times
at most, or with local operations to swap the center qubit as the star is being
generated we could bring this down to four interactions per qubit. As we see
from our generalized sequence each qubit would only need to interact twice
with the probe mode. Now if we think of a grid-like structure, the qubits
inside the graph will have $d=4$, thus using $\hat{U}$ would mean we require
8 interactions for each of these qubits. But here again if we switch to the
linear cluster sequence each one of these qubits would interact 4 times only,
twice in each of the two chains that go through it.

These multi-qubit approaches could be envisaged in different contexts.
For example we may view it as an expensive resource (be it in time or work)
alongside a cheaper one such as probabilistic two qubit parity gates.
Many schemes make use of a basic building block such as star shaped graphs.
For instance this approach could be used to directly generate the building blocks
needed in the percolation techniques derived in \cite{perco}, then easier measurement
based gates would take on from there, following the same methods. When we begin to
think of loss in the probe mode however, this star shaped graph sequence is a lot
less robust than the linear graph sequence. The reason being that halfway through
the interactions the probe holds information about all qubits, meaning that correlated
errors will be quite significant. Whereas in the linear graph sequence, the probe holds
information about at most 2 elements because it is constantly disentangling from
the previous ones. Meaning that the correlated errors will be limited to pairs of
qubits. In that sense the simplest and apparently most robust procedure for
building a grid like structure is to generate chains that would then overlap
at the intersections. In any case we can compensate for the loss in the probe
and make sure it disentangles from the qubits. This will leave us with weighted
graphs and some dephasing on the qubits. We can then resort to purification protocols
such as those proposed in \cite{purif}.

The important aspect of this measurement free approach is that it is significantly
quicker. Also it does precisely simulate the two qubit gates required for generating
the cluster state, unlike the measurement based gates which are simple parity checks.
This means that no feed forward or local operations are required. But this comes to
a price. The constraints for the strength of the interactions are greater, that is
they are now fixed, in contrast with our parity gates for which there simply was a
lower bound needed for the distinguishability of the measurement outcomes.

Now given these near deterministic operations, the number of interactions
required to build a certain cluster state becomes fixed. The question of time
then simply reduces to the number of gates we can implement in parallel.
Looking at the process in a dynamical way,
we can see now that the size of the cluster state at
a certain time during the computation is significantly reduced. This `buffer' region
of the cluster state may still be a couple of layers, but the off-line part of the
cluster, which isn't attached yet, can be made very small. Previously, the size of
the buffer that is yet to be linked up was dictated by the success probabilities of
the entangling operations \cite{bar05a}. The bigger this off-line prepared buffer is,
the more time it takes to build it and the more time it takes to attach it. In other
words the more errors it contains. Now depending on the amount of near deterministic
gates we can implement in parallel, this off-line buffer only needs to consist in a
couple of layers, freshly built, purified and attached. As a matter of fact we may not even need
this off-line aspect anymore. The individual qubits could be added directly to the
existing cluster as it is being measured. This represents huge savings in the number
of qubits we are dealing with and minimizes the error they may have picked up, as they
spend a minimal time inside a cluster state. The issues raised at the end of the last
section are still of concern here. There always will be \textit{some} waiting, between
the building and the readout, so change in support during that time, from electron spin
to nuclear spin for example, in order to minimize the dephasing, is still an important idea.

\section{Conclusion}

In this paper we have considered the usefulness of weak non-linearities in the building of
matter qubit cluster states enabling us to work in the success probability regime
of $p\geq1/2$. We first developed a 2-qubit parity check, based on a single non-linearity
per qubit and a simple measurement of the probe bus. At this point we already noticed the
advantage of using continuous variables to mediate an interaction between the qubits and
to provide us with an efficient measurement system. Then we extended the setup to a 3-qubit
entangling operation, increasing the probability of entangling a pair of qubits to 3/4. We
saw how this scheme could be generalized to using more qubits, rapidly increasing the
success probability. The 3-qubit interaction already provides new possibilities for the
schemes used in building cluster states. After what we considered the vital issues of
scaling, by going through previous results and adapting them to our own gates. The
results themselves were already significant improvements to previously proposed schemes.
The time scaling was particularly emphasized and discussed. This lead us to notice that
there will always be a compromise to be made, between the time and the number of
entangling operations required. We also observed that within this framework we
have access to measurement free approaches which can be generalized to more qubits.
They make use of the geometrical phase acquired by the probe in phase space,
simulating conditional phase gates between the qubits in the cluster. The
constraints in implementation are higher, but many scaling issues are immediately resolved.
Finally, we remark that the non-linearity and coherent state requirements for 
all these schemes to operate are satisfied by, for example, 
$\theta \sim 10^{-3}-10^{-4}$ and 
$\alpha \sim 10^{3}-10^{4}$. These realistic numbers give encouragement for experiment.

\noindent {\em Acknowledgments}: We thank J. Eisert and R. van Meter for useful discussions. This work was supported in part by
MEXT in Japan and the EU project QAP.

\end{document}